%% file: paper.tex
\documentclass[conference]{IEEEtran}
\usepackage{cite}

\usepackage[pdftex]{graphicx}
\usepackage[cmex10]{amsmath}
\usepackage{algorithmic}
\usepackage{array}
\usepackage{mdwmath}
\usepackage{mdwtab}
\usepackage{eqparbox}
\usepackage{xcolor}
\usepackage{booktabs, subcaption, tabularx}
\usepackage{url}
\usepackage{listings}
\usepackage{lstautogobble} 
\definecolor{codebgcolor}{HTML}{fdf6e3}
\begin{document}

\newcommand{\dddre}{\textsc{D\textsuperscript{3}RE}}
\newcommand{\tool}{\textsf{d3re}}
\newcommand{\ddisasm}{\textsf{ddisasm}}

\newcommand*\kris[1]{\textcolor{red}{\textbf{Kris: #1}}}
\newcommand*\yihao[1]{\textcolor{green}{\textbf{Yihao: #1}}}
\newcommand*\jeffrey[1]{\textcolor{blue}{\textbf{Jeffrey: #1}}}

\title{Declarative Demand-Driven Reverse Engineering}

\author{\IEEEauthorblockN{Yihao Sun\IEEEauthorrefmark{1},
Jeffrey Ching\IEEEauthorrefmark{2}, and Kristopher Micinski\IEEEauthorrefmark{3}}
\IEEEauthorblockA{Department of Electical Engineering and Computer Science,
Syracuse University\\
Email: \IEEEauthorrefmark{1}ysun67@syr.edu,
\IEEEauthorrefmark{2}cching01@syr.edu,
\IEEEauthorrefmark{3}kkmicins@syr.edu}}


\maketitle

\begin{abstract}
Binary reverse engineering is a challenging task because it often
necessitates reasoning using both domain-specific knowledge (e.g.,
understanding entrypoint idioms common to an ABI) and logical
inference (e.g., reconstructing interprocedural control flow). To help
perform these tasks, reverse engineers often use toolkits (such as IDA
Pro or Ghidra) that allow them to interactively explicate properties
of binaries. We argue that deductive databases serve as a natural
abstraction for interfacing between visualization-based binary
analysis tools and high-performance logical inference engines that
compute facts about binaries. In this paper, we present a vision for
the future in which reverse engineers use a visualization-based tool
to understand binaries while simultaneously querying a
logical-inference engine to perform arbitrarily-complex deductive
inference tasks. We call our vision declarative demand-driven reverse
engineering (\dddre{} for short), and sketch a formal semantics whose
goal is to mediate interaction between a logical-inference engine
(such Souffl\'e) and a reverse engineering tool. We describe a
prototype tool, \tool{}, which are using to explore the \dddre{}
vision. While still a prototype, we have used \tool{} to reimplement
several common querying tasks on binaries. Our evaluation demonstrates
that \tool{} enables both better performance and more succinct
implementation of these common RE tasks.
\end{abstract}

\input{intro}
\input{overview}
\input{design}
\input{evaluation}
\input{conclusion}

\bibliography{ref}

\end{document}

%% file: intro.tex
\section{Introduction}
\label{sec:intro}

Binary reverse engineering (henceforth RE) is the process by which we
start with some input binary (sequence of bytes) and employ various
reasoning principles to explicate its behavior when executed as
code. While RE tasks are often partially automated (e.g., via
decompilation), full automation is often impossible: the extreme
semantic expressivity afforded to binaries (including encrypted code,
stripped symbol tables, etc..) often necessitates open-ended
exploration and case-specific reasoning. Recent literature suggests
that many practicioners follow an iterative approach involving several
rounds of hypothesis formation and validation/falsification, often
assisated via a combination of static and dynamic
analysis~\cite{votipka:2020,smith:2015,johnson:2013}.

To rapidly interact with a binary, RE practicioners often use reverse
engineering tools such as Ghidra~\cite{ghidra}, IDA Pro~\cite{idapro},
or Radare2~\cite{Radare2}. The goal of these tools is to allow an
RE\footnote{When unambiguous, we will use the term RE both to mean the
  process of reverse engineering and a reverse engineering
  practicioner} to quickly explore the binary and visualize it
(typically interactively, via a GUI or CLI) in a variety of ways. For
example, a reverse engineer looking for a time bomb may first search
for calls to the system's \textsf{time} function, and then walk
backwards to understand whether each call is associated with
legitimate or malicious behavior. In doing so, the RE may need to
reason about, e.g., indirect control flow, or even identify the
\textsf{time} function (in a stripped binary). Because REs are expert
users, and often skilled programmers, RE tools provide programmatic
interfaces that enable REs to systematize reasoning tasks via
extensions. A broad range of popular extensions exist for several
tools which perform such tasks as loading the results of static
analyses~\cite{schulte:2019,gtirb}, interacting with
debuggers~\cite{retsync}, and identifying common
cryptographic-relevant code~\cite{findcrypt}.

In this paper, we argue that \emph{deductive databases} (e.g.,
Datalog) serve as a natural abstraction boundary between RE tools and
logical inference tasks over binaries. We envision a future in which a
reverse engineer interactively explores a binary using an RE tool
while simultaneously querying arbitrarily-complex logical properties
written in a terse declarative style. We call this Declarative
Demand-Driven Reverse Engineering (henceforth \dddre{}). In \dddre{},
an RE interacts with a deductive database by giving inputs (e.g., the
currently highlighted address) to a rule-based deductive inference
system written in a declarative language such as Datalog. Rules
inductively compute \emph{relations} over facts about the binary. As
an example, consider a relation \textsf{direct\_call} $\in
\textit{Addr} \times \textit{Addr}$ which relates callsite addresses
(offsets within the binary) to procedure invocation target
addresses. In our vision, \dddre{} allows REs to interactively compute
with and visualize the results of queries over these deductive rules.

We see \dddre{} as a natural extension of several observations about
the state of the start. First, many existing RE tools assemble
databases to index various properties (e.g., addresses, symbols,
etc...) of binaries for quick exploration. Deductive databases further
allow REs to write arbitrary logical queries which are computed
maximally efficiently via, e.g., compilation to relational algebra
kernels as done in Souffl\'e. Deductive databases have also enabled
several recent advances in binary analysis demonstrating both
efficiency and robustness over conventional techniques.  For example,
the Datalog-based disassembler \textsf{ddisasm} achieves both faster
and more-precise disassembly than other state-of-the-art
disassemblers, and OOAnalyzer uses Prolog to enable declarative
recovery of classes from compiled \textsf{C++} code.

In this short paper we describe our progress in implementing a
prototype tool, \tool{}, which we are building to realize the \dddre{}
vision. \tool{} allows REs to interactively define and calculate
queries of arbitrary complexity over large production binaries and
then visualize their results using Ghidra. To implement \tool{} we
have designed an interface, which we call the \emph{mediator}, that
sits between a traditional Datalog solver and an RE tool. We briefly
formalize this interaction between the RE tool and logic solver in
Section~\ref{sec:design}, and go on to describe our prototype Ghidra
extension that enables visualizing the results of binary analyses in
our tool. Using this formalism, we describe how \tool{} readily
enables a broad range of binary analyses and sketch a vision for how
we believe \dddre{} will prove to be a natural ergonomic for reverse
engineering.

We have measured the robustness of \tool{} in several ways. First, we
wanted to know whether \tool{} could truly live up to our vision of
being a natural replacement for the kinds of scripts REs already use
in their day-to-day work. To evaluate this, we reimplemented a set of
currently-existing Ghidra scripts. We happily observed that \tool{}
was not only an ergonomic advantage (allowing us to write succinct but
obviously-correct queries) but also a performance advantage. For
example, many Ghidra scripts play tricks to avoid unnecessarily
complexity that would arise in a straightforward implementation, e.g.,
iterating over a set of functions in a loop to check a property
resulting in super-linear complexity. In \tool{}, the Datalog solver
was naturally able to compile and organize work in an optimal way. We
discuss this and other results in Section~\ref{sec:evaluation}. We
conclude with a brief overview of related work and our outlook on future directions in Section~\ref{sec:conclusion}

Specifically, we claim the following three contributions:

\begin{itemize}
\item A formalization of our metadatabase as a database of databases
  used to optimize subsequent invocations of the Datalog solver
  (Section~\ref{sec:design}).

\item A prototype tool, \tool{}, consisting of a server which wraps
  \ddisasm{} with logic to enable chaining multiple subsequent calls
  via the metadatabase. Also included in \tool{} is an extension to
  the Ghidra RE toolkit to enable visualizing results computed using
  \tool{}.

\item An evaluation of \tool{} on a set of benchmarks demonstrating
  positive initial results indicating that \tool{} could replace
  present-day binary analysis infrastructure (e.g., Ghidra scripts)
  and directly enable more efficient and succinct implementation.
\end{itemize}

%% file: overview.tex
\section{Overview}
\label{sec:overview}

In this section, we demonstrate the vision and application of \dddre{}
by illustrating how a reverse engineer might explicate a vulnerability
due to an uninitialized global variable. We consider a particular
binary, \texttt{CROMU\_00038}, from DARPA's Cyber Grand Challenge
which contains a function pointer which is uninitialized when an
invalid flag is set in the metadata portion of an input
file~\cite{cgc,example}. We demonstrate how our prototype tool,
\tool{}, can be used to build a declarative query to find
uninitialized function entry points and visualize them within
Ghidra. We do not claim that \tool{} can immediately or automatically
discover vulnerabilities---in this section we try to focus on how its
declarative reasoning instead enables rapidly exploring a binary to
uncover some property.

The vulnerable segment of code is a use-before-definition bug shown in
Figure~\ref{fig:sourcecode}. The \textsf{swap\_word} function is
initialized inside of the \textsf{main} function based on a value
parsed in a TIFF header---if the flag does not match \textsf{0x4949}
or \textsf{0x4d4d} the function is left uninitialized and the call on
line 17 crashes.

\begin{figure}
\scriptsize
  \begin{lstlisting}[numbers=left]
// swap_short and swap_word only initialized within if
if (tiff_hdr->Byte_Order == 0x4949) {
  printf("Intel formatted integers\n");
  swap_word = intel_swap_word;
}
else if (tiff_hdr->Byte_Order == 0x4d4d) {
  printf("Motorola formatted integers\n");
  swap_word = motorola_swap_word;
}
#ifdef PATCHED
else {
  printf("Invalid header values\n");
  _terminate(-1);
}
#endif
// might cause an uninitialized variable bug here
offset = swap_word(tiff_hdr->Offset_to_IFD);
  \end{lstlisting}
  \caption{Uninitialized variable vulnerability in CROMU0038 source code}
  \label{fig:sourcecode}
\end{figure}

\begin{figure}
\footnotesize
  \begin{lstlisting}
>>>   load dl/use_def_global.dl
>>>   run dl/use_def_global.dl
>>>   load dl/uninitialized.dl
>>>   run dl/uninitialized.dl
>>>   highlight
>>>   comment
>>>   query use_before_def_global
00004feb	0000a180	swap_short
00005017	0000a188	swap_word
...
0000515e	0000a180	swap_short
  \end{lstlisting}
  \caption{\tool{} REPL session used in this overview.}
  \label{fig:result}
\end{figure}

\paragraph*{Loading the binary}

To begin an analysis of a binary, an RE will load the binary into a
reverse engineering tool. In our current implementation of \tool{}, a
user opens two processes silmultaneously: a GUI-based instance of
Ghidra, and a terminal running \tool{}'s REPL. The user can explore
the binary using all of the normal features of Ghidra and use all of
its conventional analyses (e.g., to recover entrypoints). However,
\tool{}'s REPL communicates with Ghidra so that when \tool{}'s
analysis finishes Ghidra's views update as appropriate.

\paragraph*{Initial processing}

It is conventional that reverse engineering tools will apply a set of
analyses to a binary to disassemble it and index various items such as
entrypoints and callsites In \tool{}, the user builds queries in
Datalog starting from a large initial set of Datalog rules that build
on top of \ddisasm{}, a Datalog-based disassembly
engine~\cite{floresmontoya:2020}. Analogously to the indexing and
analysis operations provided by Ghidra (and other RE tools), \tool{}
invokes \ddisasm{} once to build an initial database.

Building on top of \ddisasm{} was initially a strategic
choice---\ddisasm{} already includes facilities to parse object files
and transform them into input databases in the style required by
Souffl\'e. Initially, we extended \ddisasm{}'s set of rules with
additional user-specific queries---a slow process, as \ddisasm{} can
take several minutes to run on large binaries. This was at odds with
our goal of enabling rapid real-time feedback to users of \tool{}.

\dddre{} builds upon a key observation that we have found crucial to
enable efficient interactive binary analyses in practice: because
Datalog is monotonic, we can evaluate an extended program (i.e., a
program extended with a set of additional rules or queries) by using
the database resulting from the calculation of the previous
program. Thus, running \ddisasm{} \emph{once} allows pre-populating a
large set of inferred relations for a wide range of interesting facts
about binaries, including intraprocedural reachability and calling
conventions.

When a binary is loaded, \tool{} invokes \ddisasm{} with one slight
modification: every Datalog relation in \ddisasm{}'s rule database
(used by \ddisasm{} to build a disassembly) is modified to be an
output relation. In \ddisasm, only dissassembly-relevant relations are
output, rather than internal relations (e.g., those that relate to
intraprocedural reachability). By marking all \ddisasm{}'s relations
as output relations, \tool{} provides them to the user as primitives
with which to build queries over binaries\footnote{A relevant analogy
  might be that \ddisasm{} is the standard library of \tool{}}. After
the binary is loaded, all rules declared in \ddisasm{} will be
available for querying. Additionally, \ddisasm{} will be run only
once, even if the user uploads the same binary several times. All
facts generated in this step will be stored in a temporary folder on
disk managed by the metadatabase (described in
Section~\ref{sec:design}).

\begin{figure}[t]
  \begin{lstlisting}
def_global(EA,dest) :-
  code(EA), instruction_get_dest_op(EA,Index,_),
  pc_relative_operand(EA,Index,dest),
  defined_symbol(dest,_,"OBJECT","GLOBAL",_,_).
    
used_global(EA,dest,Index) :-
  code(EA), instruction_get_src_op(EA,Index,_),
  pc_relative_operand(EA,Index,dest),
  defined_symbol(dest,_,"OBJECT","GLOBAL",_,_).

def_used_global(EA_def,GA,EA_used,Index) :-
  used_global(EA_used,GA,Index),
  block_last_def_global(EA_used,EA_def,GA).

def_used_global(EA_def,GA, EA_used, Index) :-
  last_def_global(Block,EA_def,GA),
  code_in_block(EA_used, Block),
  used_global(EA_used, GA, Index),
  !block_last_def_global(EA_used,_,GA),.
\end{lstlisting}
\caption{Global Var Use-Def analysis}
\label{fig:usedef}
\end{figure}

\paragraph*{Designing a query to explicate use-before-define}

In the \dddre{} approach, REs interactively build queries to highlight
various portions of the program that match certain properties. They
then manually inspect the results of their queries and use their
intuition to build subsequent queries. Along the way, the RE may
choose to add comments to various instructions, functions, or other
forms and browse those instructions in Ghidra. In \tool{}, the
communication between the logical rules and the state of the RE tool
is reconciled by input and output tables---RE users can write queries
that consume the state of the RE tool (such as
\textsf{currentAddress}, the currently-selected address) as input
relations, perform logical inference, and leave their output in
relations such as \textsf{comment(addr,``vuln'')}.

Like \ddisasm{}, \tool{} uses the Souffl\'e Datalog engine to perform
logical inference over binaries. Users of \tool{} can incrementally
build up more rules in the interactive REPL (shown in
Figure~\ref{fig:result}). Currently, our REPL allows loading rules by
loading new files---we plan on adding direct support for new rules,
along with error-reporting feedback soon.

Knowing there was an uninitialized global function pointer being used,
a user of \tool{} might first define a set of relations to build up
def-use-chains of global variables. Datalog code to implement these
queries is illustrated in Figure~\ref{fig:usedef}. The last two rules
build up a relation
\textsf{def\_used\_global(EA\_def,GA,EA\_used,Index)}, which infers
that at address \textsf{EA\_def}, the global variable (at address
\textsf{GA}) is defined and used at address \textsf{EA\_used} at
operand index \textsf{Index}. While this is a relatively coarse query,
we envision the user could run the query on the binary to visualize a
large answer set. In our setting, this can be done using the
\textsf{highlight} or \textsf{comment} commands, which display the
data marked to be highlighted by the most recent result computation.

\begin{figure}
  \begin{lstlisting}
use_before_def_global(EA_used,GA,Name) :-
  used_global(EA_used,GA,Index),
  !def_used_global(_,GA,EA_used,_),
  defined_symbol(GA,_,"OBJECT","GLOBAL",_,Name).

use_before_def_global(EA_used,GA,Name) :-
  used_global(EA_used,GA,Index),
  def_used_global(EA_def,GA,EA_used,_),
  !def_null_global(EA_def,GA),
  defined_symbol(GA,_,"OBJECT","GLOBAL",_,Name).
  \end{lstlisting}
  \caption{uninitialized variables}
  \label{fig:uninitvar}
\end{figure}

Based on our definitions in Figure~\ref{fig:usedef}, we can define a relation
for variables which are possibly used before they are defined. We demonstrate
this in Figure~\ref{fig:uninitvar}. In the first rule, we say that if there is
some usage of a global variable at some address, but in that address, we can't
find any definition related to it, then we will consider variable there as an
uninitialized variable; The second clause says that for some usage of a global
variable even if it has some definition associated with it, if that definition
is nullptr, we will still consider that there is a use-before-def vulnerability
here.

\paragraph*{Refining the query}

In \tool{}, users can easily access the result of the rule and all
facts generated by \ddisasm{} through the GUI by writing into output
tables using \tool{} rules. Unfortunately, our above query produces
over 50 possible results---checking each occurrence would still be a
timely endeavor. Next, we narrow down the query space to the range of
just the main function. We use an auxiliary predicate,
\textsf{code\_in\_range}, which we seed with constants for the
beginning and end of the \textsf{main} function we gain from
inspecting the binary in Ghidra.

\begin{lstlisting}
code_in_range(19490,21704).
use_before_def_global(EA_used, GA, Name) :-
  code_in_range(from, to), EA_used >= from, EA_used < to,
  used_global(EA_used,GA,Index),
  !def_used_global(_,GA,EA_used,_),
  defined_symbol(GA,_,"OBJECT","GLOBAL",_,Name).
\end{lstlisting}

After new rules are applied, the output of the program becomes empty:
however, this does not specify the program is free from the
vulnerability. First, because of our constraint, only the
\textsf{main} function is searched, bugs may still hide in other
functions. Secondly, if all usage of a variable is before it’s
definition, null pointer error can still appear: programmers may
initialize a variable to \textsf{NULL} and use several non-total
branches to initialize the pointer, leaving the pointer uninitialized
at the join point when no switch fires. We modify our rules to account
for this:

\begin{lstlisting}
def_null_global(EA,GA) :-
  def_global(EA,GA), instruction_get_src_op(EA,_,Op),
  op_immediate(Op,offset), offset=0.

use_before_def_global(EA_used, GA, Name) :-
  code_in_range(from,to), EA_used >= from, EA_used < to,
   used_global(EA_used,GA,Index),
   def_used_global(EA_def,GA,EA_used,_),
   !def_null_global(EA_def,GA),
   defined_symbol(GA,_,"OBJECT","GLOBAL",_,Name).
\end{lstlisting}

\begin{figure}[b]
  \centering \includegraphics[width=.45\textwidth]{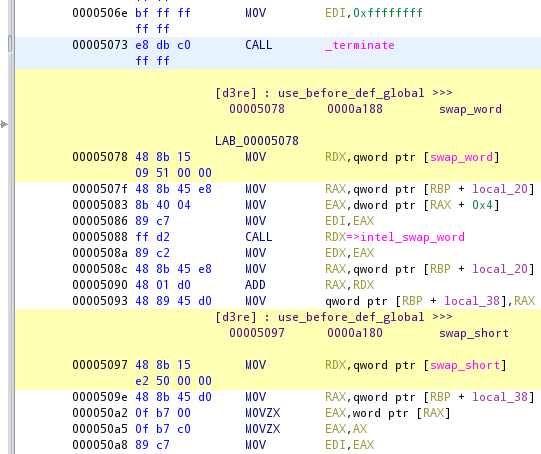}
  \caption{Ghidra with highlights and comments declaratively specified
    to output results inferred via \tool{} for our example.}
  \label{fig:codeview}
\end{figure}

This change results in 19 addresses to search, and combining these
results with use-def information in the previous step and
intra-procedural control-flow graph in Ghidra, we can fairly easily
infer that the global variable \texttt{swap\_word} is initialized to
\textsf{0} at address \texttt{0x4c2a}, that both conditional jumps
\texttt{0x4f80} and \texttt{0x4fb8} fail, and observe a subsequent
usage of \texttt{swap\_word} at \texttt{0x5017} which will trigger a
crash. At any stage in our process, we can sync Ghidra's UI with the
current database using several REPL commands (an example is shown in
Figure~\ref{fig:codeview}). In a fully-fledged implementation of
\tool{}, we hope to have UI gadgets (or templates) to help users
interactively build queries. For example, we may allow the user to
select a region of the binary and build a rule that applies only to
that region, or right-click on a function and build a rule specific to
callers of that function. We believe this will need to be informed by
a combination of interviews with expert users, participatory design,
and (perhaps) user studies. This is work we plan to undertake now that
we have proven initial success to ourselves with \tool{}.

We conclude this section by remarking upon the nature of our
analyses. Our analyses would be considered na\"ive by the standards of
industrial static analyses. Indeed, our reasoning is not even
sound---we can restrict ourselves to looking at results for only one
function or ignore complex behavior. Still, we believe that this
iterative ad-hoc reasoning is a technique many reverse engineers
already employ---the vision of \dddre{} is to harmoniously leverage
state-of-the-art deductive reasoning engines while performing
human-guided RE tasks.

%% file: design.tex
\section{Design and Implementation}
\label{sec:design}

\begin{figure}[t]
  \centering
  \includegraphics[width=.45\textwidth]{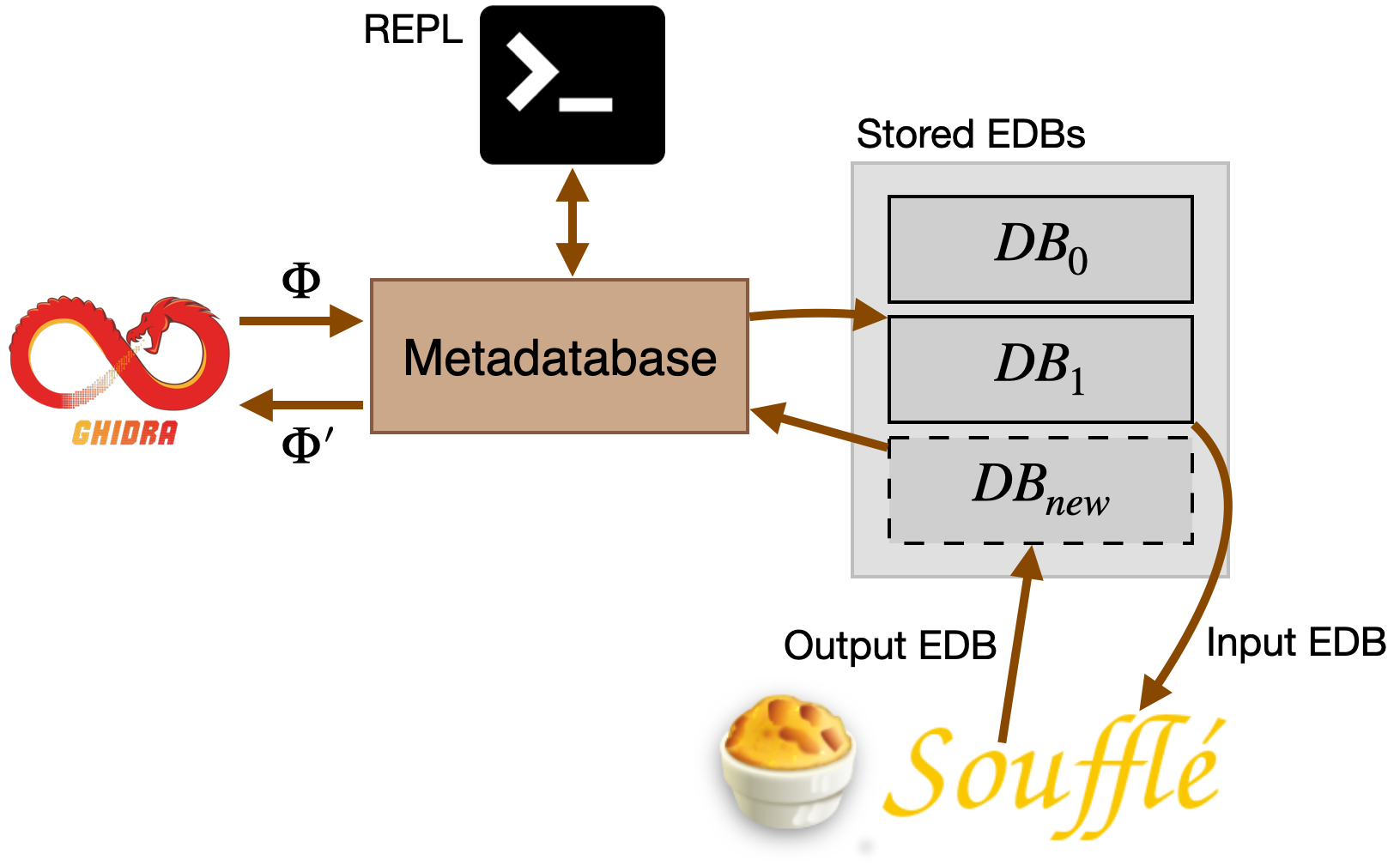}
  \caption{High-level components and their interactions in \tool{}.}
  \label{fig:design}
\end{figure}

In this section, we present both a formal semantics for \dddre{} and
describe our implementation of \tool{}. The high-level architecture of
\tool{} is outlined in Figure~\ref{fig:design}. Conceptually, the key
idea of our semantics is to maintain a \emph{metadatabase} to allow
efficient incremental reuse of previously-computed databases. In
\tool{}, this metadatabase takes the form of a server which accepts
Datalog programs to run to a fixed-point.

The metadatabase (server) interacts with both the REPL process and
Ghidra to render output databases into view annotations (e.g.,
highlights or comments) in the Ghidra UI based on REPL commands. Our
visualization is currently limited to printing to Ghidra's console,
highlighting a set of lines (typically some output relation), or
annotating a line with a comment (whose contents may be dynamically
determined via a Datalog query). We plan to investigate adding
comments to other Ghidra UI elements (such as inferred classes) and
other visual integration as future work.

\subsection{Formal semantics of \dddre{}}
\label{sec:formal}

Due to space restrictions, we present only a sketch of a formal
semantics for \dddre{}. Semantics of Datalog programs are typically
phrased in terms of an extensional database (EDB), an
extensionally-enumerated set of ground facts, and intensional database
(IDB), the set of rules defining the program~\cite{datalog}. Datalog's
semantics is given by a least-fixed-point of an ``immediate
consequence'' operator over the rules for the program. Because Datalog
programs have a finite Herbrand base (sets of atoms), this fixed-point
necessarily exists (though in practice Datalog engines allow
extra-logical behavior such as arithmetic). Datalog's conventional
semantics is \emph{monotonic}, in the sense that strictly more facts
are accumulated as the fixed-point computation evolves---negation is
allowed only when it may be stratified.

We define an \emph{EDB metadatabase} as a graph of EDBs with labeled
edges, $(\Delta, \stackrel{P}{\rightarrow})$, where $\Delta$ is a set
of EDBs, each EDB enumerating tuples for a given set of relations, and
$\stackrel{P}{\rightarrow}$ is a relation in $\Delta \times
\textit{Rules} \times \Delta$. When we process a program, $P$, using
an input EDB, we traverse the graph $(\Delta,
\stackrel{P}{\rightarrow})$ to find the most optimal,
\emph{compatible} EDB to start execution of $P$. Aided by Datalog's
monotonicity, we define an EDB as compatible if it was produced by a
subset of rules (or facts) from the input program / EDB. We conclude
our formalism sketch by remarking that $(\Delta,
\stackrel{P}{\rightarrow})$, given our usage, also forms a lattice.

\subsection{Implementation of \tool{}}

\tool{} is implemented in two parts: a REPL that communicates with
Ghidra's GUI and a background service to manage the metadatabase and
run the Datalog engine. The REPL currently communicates with Ghidra
via a third-party extension named
\textsf{ghidra\_bridge}~\cite{bridge}, which we plan to replace
imminently with an extension using protocol buffers.

To execute a Datalog program P, \tool{} analyzes the file using the
logic sketched in the above section to determine an optimal compatible
EDB to use. In the common case, a user will gradually accumulate a
stream of programs $P$, $P'$, $P'''$ consisting of a mix of rules and
assumptions. In the future, we envision that certain assumptions
(e.g., about calling conventions) may be implemented as GUI extensions
rather than, e.g., manually-enumerated facts. After each run, the
metadatabase will index the output facts and associate them with the
program $P$, establishing an edge in the aforementioned graph. In our
experiments, we refer to this as ``caching.''

%% file: evaluation.tex
\section{Evaluation}
\label{sec:evaluation}

\newcolumntype{n}{X}
\newcolumntype{s}{>{\centering\hsize=.4\hsize}X}
\newcolumntype{t}{>{\centering\hsize=.5\hsize}X}
\newcolumntype{l}{>{\centering\hsize=1\hsize}X}

\begin{table}[t]
  \centering
  \normalsize
  \caption{Script size (lines of code) of Ghidra script (Python) vs. \tool{} Datalog}
  \label{tab:loc}
  \begin{tabularx}{.45\textwidth}{@{}nnn@{}} \toprule
                 & {Ghidra Python} & {\tool{} Datalog} \\ \midrule
    {non-xor}    & 33              & 8                 \\
    {overflow}   & 60              & 18                \\
    {basicblk}   & 37              & 4                 \\
    {findcrypto} & 166             & 45                \\ \bottomrule
  \end{tabularx}
\end{table}

\begin{table}[t]
  \footnotesize
  \centering
  \caption{Running time of Ghidra scripts vs. equivalent
    implementation in \tool{} (all numbers in seconds).}
  \label{tab:ghidravsdddre}
  \begin{tabularx} {.5\textwidth}{@{}nssssss@{}} \toprule
              & {bison} & {souffle} & {gzip} & {re2c} & {redis} & {rsync} \\ \midrule
  {non-xor Ghidra}               & 3.569 & 107.5         & 2.205 & 3.903 &
  10.52 & 3.050 \\
  {non-xor \tool{}}                   & 0.518 & 6.515         & 0.097 & 0.756 &
  1.306 & 0.486  \\ \midrule
  {overflow Ghidra} & 0.370 & 0.247         & 0.600 & 0.240 &
  0.760 & 0.180 \\
  {overflow \tool{}}     & 0.617 & 0.319         & 0.051 & 0.094 &
  0.095 & 0.044 \\ \midrule
  {basicblk Ghidra}           & 340.6 & {--} & 4.664 & 472.1 &
  1806  & 107.4 \\
  {basicblk \tool{}}               & 0.539 & 7.13          & 0.094 & 0.812 &
  1.433 & 0.571 \\ \midrule
  {findcrypt Ghidra}  & 0.207 & 1.033 & 0.224 & 0.214 & 0.475 & 0.289 \\
  {findcrypt \tool{}} & 1.287 & 14.53 & 0.224 & 1.701 & 2.938 & 1.186 \\ 
  \bottomrule
  \end{tabularx}
\end{table}

\begin{table}[t]
  \footnotesize
  \centering
\caption{Runtime of successive invocations to \tool{} with (C) and without (S) rule caching.}
\label{tab:cached}
\begin{tabularx} {.5\textwidth}{@{}nlllll@{}} \toprule
                    & {ddisasm} & {stack\_var} & {heap\_var} & {static\_var} & {unl\_static} \\ \midrule
  {\textsf{souffle} C}  & 170       & 11.88       & 58.35      & 5.008        & 0.039              \\
  {\textsf{souffle} S}     & 170       & 11.79       & 66.02      & 67.00        & 66.52              \\ \midrule
  {\textsf{bison} C}    & 7         & 0.932       & 1.409      & 0.545        & 0.022              \\
  {\textsf{bison} S}       & 7         & 0.934       & 1.916      & 2.122        & 2.075              \\ \midrule
  {\textsf{re2c} C}     & 9         & 1.457       & 4.417      & 0.704        & 0.025              \\
  {\textsf{re2c} S}        & 9         & 1.494       & 5.257      & 5.449        & 5.458              \\ \midrule
  {\textsf{redis} C}    & 11        & 1.918       & 2.544      & 1.302        & 0.025              \\
  {\textsf{redis} S}       & 11        & 1.919       & 3.525      & 3.712        & 3.726              \\ \midrule
  {\textsf{rsync} C}    & 8         & 0.766       & 0.908      & 0.481        & 0.028              \\
  {\textsf{rsync} S}       & 8         & 0.783       & 1.325      & 1.423        & 1.384              \\ \bottomrule
\end{tabularx}
\end{table}

We evaluated \tool{} both qualitatively, by implementing several
queries, and quantitatively by measuring its performance in
benchmarks. While \tool{} is still a work in progress, we had several
hypotheses we aimed to test as we designed and conducted these
experiments. First, we wanted to understand whether \tool{} provided
the necessary building blocks to enable replacing currently-existing
Ghidra scripts. Second, we wanted to understand whether \tool{} could
offer performance competitive with the kinds of Ghidra scripts that
reverse engineers typically use. Last, we wanted to understand the
performance of Ghidra for performing several repeated queries that
might mirror a realistic end-to-end workload using \tool{}.

\paragraph*{Ghidra Script Replication Study}

we wanted to determine whether \tool{} could realistically be used to
accomplish the kinds of tasks that reverse engineers face on a
day-to-day basis. This is an admittedly challenging question, which we
plan to eventually evaluate in several ways including user
studies. However, as initial work in this direction we arbitrarily
selected four Ghidra scripts listed in the \texttt{awesome-ghidra}
GitHub repository\cite{awesomeghidra}. The scripts we chose are listed
in Table~\ref{tab:loc}, along with their corresponding lines of code
in Python / Datalog. While Ghidra scripts may consist of a mix of
Python and Java, our experience is that most scripts use a small
subset of the Python API. The first three are relatively small and
find instructions that match a specific template, e.g.,
\textsf{non-xor} finds \textsf{xor} instructions that aren't zeroing
registers, and \textsf{overflow} heuristically searches for potential
overflows in calls to common functions such as \textsf{strcpy}. Our
largest was \textsf{findcrypto}, which looks for common cryptographic
constants.

\paragraph*{Qualitative Results of our Replication Study}

Our experience using \tool{} to replace Ghidra scripts must be
understood in the context that we are expert users and the developers
of \tool{}. However, we are pleasantly surprised that \tool{} enabled
us to succinctly write equivalent implementations of each Ghidra
script: we rewrote each script in substantially less Datalog
code. This is because the declarative nature of Datalog eliminates the
need for much of the conventional ceremony around, e.g., looping over
instructions and checking against a type that we found in our
evaluation scripts. Key to \dddre{}'s success, we believe, is its
ability to directly use relations from \ddisasm{}: we found that much
of the necessary work of, e.g., filtering instructions by their type
or operand was very useful at achieving succinct Datalog in
practice. We are in the initial planning stages of developing a
reverse engineering tutorial (or mini-course) around \tool{}, and are
hoping to use this to recruit developers to get more realistic
assessment of \tool{}'s usability by professional REs.

\paragraph*{Quantitative Results of our Replication Study}

We hoped that \tool{}, being based on a high-performance Datalog solver, would
offer performance competitive with Ghidra's current scripts. Each of our
evaluation scripts processed the entire binary and would highlight or label
certain instructions. To test the Ghidra scripts, we used Python's standard
\textsf{time} function before and after the script's work finished. We evaluated
the corresponding Datalog program by using Souffl\'e's internal performance
timers. We then benchmark Ghidra vs. \tool{} on a corpus of six binaries (all
sized less than 10MB), five from \ddisasm{}'s test suite and Souffl\'e, shown at
the top of Table~\ref{tab:ghidravsdddre}. We used the latest versions of each
pre-built in the latest Arch Linux, but we used a pre-built version of
Souffl\'e. For each script, we waited for all of Ghidra's typical analyses to
finish, and similarly we ran \ddisasm{} to build up the initial input database
for \tool{}.

The body of Table~\ref{tab:ghidravsdddre} compares the runtime of each
Ghidra script versus its corresponding implementation in \tool{}. The
single occurrence of -- indicates that Ghidra did not finish within an
hour. Broadly, we found that \tool{} outperformed Ghidra for each of
the scripts in our replication study. As we had hoped, \tool{}'s
design allowed us to leverage useful relations from \ddisasm{}. We
found that many scripts do things like naive loops over sets of
functions or symbols to locate some property. By contrast, the
declarative style of \tool{} allowed us to write these not only more
succinctly (e.g., Datalog naturally aggregates results) but also more
efficiently---Souffl\'e optimally compiles input programs to efficient
relational algebra kernels that loop only when necessary. We did
observe various ways in which \tool{}'s limitations could cause
performance issues. For example, the \textsf{findcrypto} script scans
the binary for 256-segments of code. \tool{} is built on Souffl\'e,
which supports 64-bit primitive ints, but not 256-byte sequences. Thus,
we had to build up sequences via a set of Datalog rules, causing
inefficient memory representation due to the necessary duplication due
to representing subsequences as Datalog facts.

\paragraph*{Evaluating End-to-End Behavior in Subsequent Invocations}

To understand the effect of caching via repeated calls to \tool{}, we
ran four subsequent analysis queries in a row using both our
caching-based approach and without caching (wherein we started only
with the results of \ddisasm{}). Our results are shown in
Table~\ref{tab:cached}: the time of the cached run (C) is shown above
the time for the correspond sequential run (S). As each query builds
on the previous, we expect caching to reduce the amount of work and
commensurately reduce the runtime. \text{stack\_var} finds
stack-allocated variables, while \textsf{heap\_var} calculates stack
variables holding pointers to heap values based on
\textsf{stack\_var}. \textsf{static\_var} and \textsf{unl\_static}
attempt to find uninitialized global variables. Overall, we found rule
caching was especially important on larger binaries versus sequential
runs, justifying our choice to structure the metadatabase as a graph.

%% file: conclusion.tex
\section{Related Work and Conclusion}
\label{sec:conclusion}

We conclude with a brief discussion of proximately-related work that
lies at the synthesis of reverse engineering and static / dynamic
analysis, and contextualize this work in terms of our aspirations for
the future of \dddre{}. There has been extensive work using logic
programming, and in particular Datalog, for static analysis of
higher-level langauges such as
Java~\cite{smaragdakis:2014,jordan:2016,scholz:2016}. The success of
the Souffl\'e Datalog engine has inspired recent adoption of logic
programming within the binary analysis community. For example, Datalog
Disassembly uses Souffl\'e to achieve both faster and more-precise
disassembly than the state-of-the-art disassembler
Ramblr~\cite{floresmontoya:2020}. Similarly, OOAnalyzer uses
XSB-Prolog, a version of Prolog implemented as a
library~\cite{schwartz:2018}. We are currently reimplementing
OOAnalyzer in \tool{} targeting \textsf{C++} Linux binaries. We feel
particularly excited about this direction because we believe Souffl\'e
will be immediately more scalable than XSB-Prolog.

While there are a broad range of plugins for Ghidra and IDA Pro to
load the \emph{results} of static analyses, we believe \tool{} is the
first to focus on the combination of open-ended deductive logical
inference and rapid interactivity (enabled by our metadatabase). We
believe the most closely related work is Ponce~\cite{ponce}, which
enables GUI-based symbolic execution. We plan to integrate symbolic
execution into \tool{} as a long-term goal, inspired by the recent
work of Formulog~\cite{formulog}.

Our goal in this work was to introduce a new vision for reverse
engineering, \dddre{}, wherein expert users rapidly query
high-performance logical inference engines to help them accomplish
their day-to-day work in RE, vulnerability construction, and
penetration testing. Visualization-based tools such as Ghidra are of
immense value in understanding a binary, but have fundamentally
different design considerations than high-performance logical
inference enginges (such as Souffl\'e). Recent work in compiling
Datalog to parallel relational algebra (e.g., Gilray et
al.~\cite{hipc}) has enabled a new frontier in scale of Datalog-based
analyses. We hope that developments such as these will someday enable
realizing fully the vision of \dddre{} to help reverse engineers
perform powerful static binary analyses at unprecedented scale.